# An investigation of the two-dimensional non-Hermitian Su-Schrieffer-Heeger Model

Udai Prakash Tyagi and Partha Goswami

*Physics Dept., D.B. College (University of Delhi), Kalkaji, New Delhi- 110019, India*

*Email of the first author: uptyagi@yahoo.co.in*

*Email of the corresponding author: physicsgoswami@gmail.com*

## Abstract

This communication presents an examination of a two-dimensional, non-Hermitian Su-Schrieffer-Heeger (SSH) model, which is differentiated from its conventional Hermitian counterpart by incorporating gain and/or loss terms, mathematically represented by imaginary on-site potentials. The time-reversal symmetry is disrupted due to these on-site potentials. Exceptional points in a non-Hermitian system feature eigenvalue coalescence and non-trivial eigenvector degeneracies. Utilization of the rank-nullity theorem and graphical analysis of the phase rigidity factor enable identification of true exceptional points. Furthermore, this investigation achieves vectorized Zak phase quantization and examines a topolectric RLC circuit to derive the corresponding topological boundary resonance condition and the quantum Hall susceptance. Although Chern number quantization is not feasible, staggered hopping amplitudes corresponding to unit-cell lattice sites lead to broken inversion symmetry with non-zero Berry curvature, resulting in finite anomalous Nernst conductivity.

### 1.Introduction

The two-dimensional non-Hermitian Su-Schrieffer-Heeger (SSH) model Hamiltonian **[1,2]**, representing a generalized form of the one-dimensional standard Hermitian SSH model **[3-5]**, is used to describe topological phases in systems exhibiting non-Hermitian dynamics. To investigate a 2D model, which is the aim of this paper, it is essential to first mention a few fundamental facts related to the 1-D Hermitian model **[3]** and then examine the effects of incorporating an additional dimension and non-Hermitian terms. The 1D model was initially developed to describe the electronic properties of polyacetylene, a conjugated polymer chain with two different sublattices. The different sublattices A and B involve the hopping parameters $v$ and $w$ between them. The reciprocal-space Hamiltonian matrix corresponding to the Hermitian version of 1D may be represented as $H_{SSH}(k) = (v + w\cos(ak)\sigma_x + (-w\sin(ak))\sigma_y]$ where $\sigma_j$' s are Pauli matrices. As long as the hopping amplitudes are staggered ($v \neq w$), there is an energy gap separating the lower, filled band, from the upper, empty band. However, when $v = w$, the SSH model describes a conductor. A key fact in this context is the presence of topological edge states when the system is in a topologically nontrivial phase. In contrast, the non-Hermitian one-dimensional SSH models **[4,5]** lead to phenomena such as complex energy spectra, dissipation, and exceptional points (EPs) **[6-12]**.

This paper examines a non-Hermitian two-dimensional Su-Schrieffer-Heeger (2DSSH) model, differing from its Hermitian counterpart by incorporating gain and/or loss terms, which disrupt time-reversal symmetry (TRS) *T*. The parity-time (*PT*) symmetry and the particle-hole symmetry (PHS), however, remains unbroken under certain conditions (see section 2). In Figure 1 (a), a pictorial representation of this two-dimensional model has been shown. Here,

A, B, C, and D correspond to unit-cell sites of the model (square) lattice with lattice constant 'a'. While the symbols $u$ and $t_1$ stand for the hopping parameters along $x$- direction (horizontal), the symbols $v$ and $t_2$ are those along $y$- direction (vertical). The formulation of this Hamiltonian is presented below in Eq.(1) in real space. The imaginary staggered potentials (ISPs) are to be introduced as supplementary terms in the reciprocal space representation $H_{NH,2D}(\mathbf{k})$ in Eq.(5). These potentials, representing non-Hermitian strengths within the unit cell, are assigned values of $(i\gamma, -i\gamma)$ and $(i\gamma, -i\gamma)$ for (A, B) and (C, D), respectively. The reason for their inclusion is that gain/ loss may not be insignificant in real systems. We choose the wavenumber $\mathbf{k} = (k_x, k_y)$ to assume values within the first Brillouin zone (BZ) in the Hamiltonian $H_{NH,2D}(\mathbf{k})$. It will be shown in Section 2 that the Hamiltonian $H_{NH,2D}(k_x, k_y)$ is PT-symmetric [13] if $u = t_1$. Conversely, for $u \neq t_1$, the Hamiltonian exhibits complex eigenvalues, marking the onset of a broken PT-symmetric phase. A discussion on the symmetries associated with $H_{NH,2D}(k_x, k_y)$ is provided in Section 2.

An important aspect of non-hermiticity of the Hamiltonian is that it exhibits EPs [6-12] as mentioned above. EPs in non-Hermitian Hamiltonians are associated with the merging of multiple eigenvalues and their corresponding eigenvectors, often causing the eigenvectors to become linearly dependent. This coalescence frequently leads to a non-Hermitian Hamiltonian lacking a complete set of orthonormal eigenvectors, necessitating the interpretation of orthogonality in a generalized sense involving a biorthogonal basis. This blurs the distinction between 'state' and 'observable'. To revive a probabilistic interpretation, a novel inner product is introduced, leveraging a non-trivial metric operator η (see also section 5). This operator generally possesses off-diagonal elements, indicating that it introduces long-range correlations between otherwise spatially separated components, which reflects the delocalized nature of the biorthogonal basis in the non-Hermitian regime. In particular, non-Hermitian systems introduce left and right eigenvectors as a generalized form of orthogonality [14-16], where $H_{NH}|u^{(j)}\rangle = E_j|u^{(j)}\rangle$ (right eigenvectors), and $H_{NH}^\dagger|v^{(j)}\rangle = E_j^*|v^{(j)}\rangle$ or, $\langle v^{(j)}|H_{NH} = E_j\langle v^{(j)}|$ (left eigenvectors). Here, the eigenvalues $E_j$ and the corresponding eigenvectors ($|u^{(j)}\rangle, \langle v^{(j)}|$) of the Hamiltonian in Eq.(5) are indexed by $j = 1,2,...$. Moreover, these eigenvectors are not individually orthonormal in the standard sense, but collectively form a bi-orthonormal system satisfying $\langle v^{(i)}|u^{(j)}\rangle = \delta_{ij}$. The Gram matrices for the right and left eigenvectors are not generally inverses of each other, i.e., $(\langle u^{(i)}|u^{(j)}\rangle) \neq (\langle v^{(i)}|v^{(j)}\rangle)^{-1}$, but the Gram matrices built from right/left eigenvectors can be inverses if the eigenvectors are bi-orthonormalized. Additionally, an indicator of true EPs is the phase rigidity given by the relation $P_j = |\langle v^{(j)}|u^{(j)}\rangle|/|\langle u^{(j)}|u^{(j)}\rangle|$. While a Hermitian system has $P_j = 1$ for all $j$, when approaching an EP, $P_j \to 0$ for the states that coalesce in a non-Hermitian system. Therefore, at EPs, one may focus on this indicator. We report the outcome of our investigation of this issue in Section 2 of this paper. Furthermore, the engineering of topologically insulating and conducting phases in non-Hermitian systems is enabled by manipulating their gain and loss terms. We also show in Section 2 that our non-Hermitian model provides a versatile framework for creating insulating and conducting phases by controlling the interplay of gain (amplification) and loss (attenuation).

It is pertinent to mention that the Zak phase [17,18] is a concept from topological phases of matter, typically defined for one-dimensional (1D) periodic systems, such as $H_{SSH}$. It quantifies the geometric phase that arises when a wavefunction is adiabatically transported around the Brillouin zone, often used to describe systems with topological characteristics. This phase, which is a topological invariant of $H_{SSH}$ like 1D-models, predicts the existence or absence of edge states in various cases [18]. The phase is measured in modulo ($2\pi$). Thus, the phase is quantized to $n\pi$ where $n = 0,1$. For our 2D system, we have obtained the vectorized Zak phase components $\phi_x$ and $\phi_y$ as a function of $\frac{u}{v}$ in Section 3 for both $\gamma = 0$ and $\gamma \neq 0$. The plots in Figure 4(a) and 4(b) for $\gamma = 0$ indicate that the conventional bulk-boundary correspondence (BBC) depends crucially on the ratio $\frac{u}{v}$. In non-Hermitian systems ($\gamma \neq 0$), exceptional points and broken symmetries can complicate causing edge modes to lose robustness or abrogate BBC. To resolve this, bi-orthonormal inner products have been employed in the analysis here without disrupting PT symmetry. This facilitates precise predictions of edge states as shown in Figure 4(c) and 4(d).

In this communication, we also consider an 2D RLC topolectric circuit [19, 20] arranged in a lattice-like structure as shown in Figure 1(b), where the resistors (R), inductors (L) and capacitors (C) form a periodic structure. i.e. the one with repeating unit cells. The circuit's dynamics are governed by Kirchhoff's laws, and the admittance (or impedance) matrix (which is a function of frequency) of the circuit can describe the system's response. The equations of motion for the voltages across capacitors and the currents through inductors can be written in a matrix form, where the Hamiltonian is related to the impedance of the circuit. One needs to solve for the eigenstates of the circuit's Hamiltonian matrix. This corresponds to finding the modes of the circuit's oscillations. The eigenstates are typically periodic wavefunctions in frequency space. The quantum Hall susceptance is calculated utilizing the eigenfunctions corresponding to the Laplacian matrix in Eq.(13) below. It is found to be positive which indicates that the system has capacitive properties.

The quantum anomalous Hall (QAH) state is a topologically nontrivial phase characterized by an integer value of the Chern number and the presence of robust edge states that are dissipation-less. The QAH effect typically occurs in systems with strong spin-orbit coupling and in the presence of a ferromagnetic order that breaks TRS. In a system with time reversal symmetry (TRS) (and inversion symmetry (IS)), the Chern number must be zero because TRS requires that for every state at momentum $k$, there is a time-reversed state with the opposite contribution to the Hall conductivity. The SSH model [3-5], in its standard form, does not naturally support the quantum anomalous Hall state with an integer Chern number, for the standard SSH model with real hopping terms preserves TRS. However, it is possible to modify the model to include mechanisms that break TRS introducing a magnetic field or exchange interaction. This would break TRS and induce a nonzero Chern number, leading to a topologically nontrivial phase with edge states. We shall show in section 4 that the introduction of non-hermiticity, leading to broken TRS, does not inherently create a QAHE with an integer Chern number. Similarly, the staggered hopping amplitudes, leading to the broken IS, do not necessarily create such a QAHE. The topological properties of the system are further highlighted by the quantum anomalous Nernst effect (QANE), a phenomenon in which a thermoelectric current, such as the Nernst effect, is induced by an anomalous transverse thermoelectric response in a system with nonzero Berry curvature and broken IS. We will show in Section 4 that QANE is possible for the present system, as broken IS gives rise to a nonzero Berry curvature and nonzero, albeit non-integer, Chern number in certain parameter windows. Additionally, while QAHE is contributed by the sum of Berry curvatures of all occupied bands, QANE is computed by the Berry curvature close to the Fermi level. This implies that an insubstantial or substantial anomalous Hall effect does not necessarily imply an inconsiderable or considerable anomalous Nernst effect, due to differences in their underlying mechanisms.

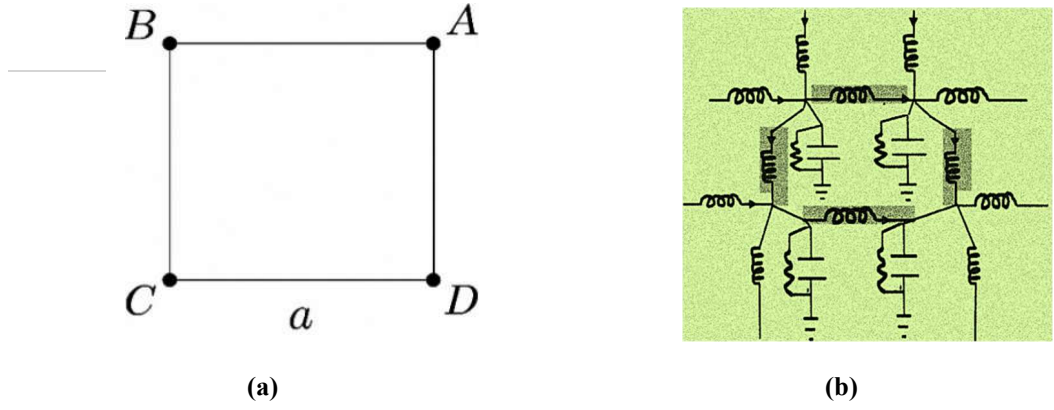

(a) (b)

**Figure 1.** (a) A pictorial representation of the two-dimensional model square lattice, where A, B, C, and D correspond to unit-cell sites of the model (square) lattice with lattice constant '$a$', starting with A in the upper right corner, B in the upper left corner, C in the lower left corner and D in the lower right corner. Here, $u$ ($t_1$) stands for the hopping parameter along horizontal (x-) direction joining A and B (C and D). Similarly, $v$ ($t_2$) stands for the hopping parameters along vertical (y-)direction joining B and C (D and A). The imaginary staggered potentials applied to sites (A, B) and (C, D) are ($i\gamma, -i\gamma$) and ($-i\gamma, i\gamma$), respectively. (b) A topolectric RLC circuit, with a four-node (represented by blue (B), red (R), green (G), and yellow (Y) colours) unit cell, is shown. The dark bases are for inductors $L_2$, while the remaining inductors are $L_1$. Node voltages $V_1$, $V_2$, $V_3$, and $V_4$ correspond to sites marked by blue, red, green, and yellow colours, respectively. Resistors R are connected in parallel with capacitors C within the cell. Moreover, currents $I_1$ and $I_2$ are fed into nodes B and Y, respectively, from the left-hand side, and currents $I_3$ and $I_4$ are fed into nodes B and R, respectively, from the top. Additionally, currents $I_5$, $I_6$, $I_7$, and $I_8$ flow through the four arms of the unit cell. Specifically, currents $I_5$ and $I_6$ flow from the top horizontal arm and the bottom one, respectively. Similarly, currents $I_7$ and $I_8$ flow from the left and right vertical arms, respectively.

The paper is organized as follows: We present the 2D SSH model in section 2 and discuss its symmetry properties. We obtain the energy eigenvalues and the corresponding eigenvectors in this section in PT symmetric/ non-symmetric cases. We calculate the Zak phase in Section 3 leading to the bulk-boundary correspondence. In this Section, we also investigate a topolectric RLC circuit as an application of the model presented. We present a discussion in Section 4 relating to a case where the hopping amplitudes are staggered in $\hat{y}$ - direction. The paper ends with discussion and outlook in section 5.

## 2. Model and Method

### Model

The 2D model lattice considered here comprises of sites located at $\{\mathbf{R}, \mathbf{R} + \Delta R_x \hat{x}, \mathbf{R} + 2\Delta R_x \hat{x}, \ldots\}$ in the x-direction and same number of sites located at $\{\mathbf{R}, \mathbf{R} + \Delta R_y \hat{y}, \mathbf{R} + 2\Delta R_y \hat{y}, \ldots\}$ in the y - direction. In the absence of ISPs, the 2D model can be described by the following the effective tight binding model:

$$H_{NH,2D} =$$

$$\sum_R \{u \ |\mathbf{R}, A\rangle\langle \mathbf{R}, B| + t_1 \ |\mathbf{R} + \Delta R_x \hat{x}, B\rangle\langle \mathbf{R}, A| + t_1 \ |\mathbf{R}, C\rangle \langle \mathbf{R}, D| + u \ |\mathbf{R} + \Delta R_x \hat{x}, D\rangle\langle \mathbf{R}, C| + v \ |\mathbf{R}, B\rangle \langle \mathbf{R}, C| + t_2 \ |\mathbf{R} + \Delta R_y \hat{y}, C\rangle \langle \mathbf{R}, B| + t_2 \ |\mathbf{R}, D\rangle \langle \mathbf{R}, A| + v \ |\mathbf{R} + \Delta R_y \hat{y}, A\rangle \langle \mathbf{R}, D|\} + h.c..$$  (1)

The effective tight binding model of a crystal, in fact, is a theoretical model that describes the behavior of the crystal's vibrations in terms of the interaction between its constituent atoms or molecules. In this model, the crystal is treated as a lattice of discrete points, or "sites," each of which represents an atomic or molecular unit. The strength of the interaction between neighboring sites is characterized by a set of parameters, such as the hopping energies. These are $\{u, v, t_1, t_2\}$ in the present problem. These parameters are typically determined by fitting the model to experimental data or more detailed calculations, and can be used to obtain the energy dispersion relations.

We will now write down the Hamiltonian introduced in Eq.(1) in momentum space. As the first step, we note that the Hamiltonian, in the absence of ISP, may also be written slightly differently in terms of the on-site creation operator $\{a_{R,1}^\dagger, b_{R,2}^\dagger, ...\}$ and the on-site annihilation operators $\{a_{R,1}, b_{R,2},....\}$ as

$$H_{NH,2D} = \sum_R [u a_{R,1}^\dagger b_{R,2} + u b_{R,2}^\dagger a_{R,1} + t_1 c_{R,3}^\dagger d_{R,4} + t_1 d_{R,4}^\dagger c_{R,3} + v b_{R,2}^\dagger c_{R,3} + v c_{R,3}^\dagger b_{R,2} +$$

$$t_2 d_{R,4}^\dagger a_{R,1} + t_2 a_{R,1}^\dagger d_{R,4} + u d_{R,4}^\dagger c_{R+\Delta R_x \hat{x},3} + u c_{R+\Delta R_x \hat{x},3}^\dagger d_{R,4} + v d_{R+\Delta R_y \hat{y},4}^\dagger a_{R,1} +$$

$$v a_{R,1}^\dagger d_{R+\Delta R_y \hat{y},4} + t_1 b_{R+\Delta R_x \hat{x},2}^\dagger a_{R,1} + t_1 a_{R,1}^\dagger b_{R+\Delta R_x \hat{x},2} + t_2 b_{R,2}^\dagger c_{R+\Delta R_y \hat{y},3} +$$

$$t_2 c_{R+\Delta R_y \hat{y},3}^\dagger b_{R,2}]. \quad (2)$$

We now apply periodic boundary conditions $|R + N\Delta R, P\rangle = |R, P\rangle$ in the x- and y-directions, where N is the number of sites in these directions. The Fourier transform $|R_x, P\rangle = N^{-\frac{1}{2}} \sum_k e^{ik_x R_x} |k, P\rangle$ and $|R_y, P\rangle = N^{-\frac{1}{2}} \sum_k e^{ik_y R_y} |k, P\rangle$ eventually yields $H_{NH,2D} =$

$$\sum_R \sum_k \{[\frac{u}{N} |k,A\rangle\langle k,B| + \frac{t_1}{N} \exp(ik_x(R_x + \Delta R_x))|k,B\rangle\langle k,A| \exp(-ik_x R_x) +$$

$$h.c.] + \sum_r \sum_k \{[\frac{t_1}{N} |k,C\rangle\langle k,D| + \frac{u}{N} \exp(ik_x(R_x + \Delta R_x))|k,D\rangle\langle k,C| \exp(-ik_x R_x) +$$

$$h.c.]\} + \sum_R \sum_k \{[\frac{v}{N} |k,B\rangle\langle k,C| + \frac{t_2}{N} \exp(ik_y(R_y + \Delta R_y))|k,C\rangle\langle k,B| \exp(-ik_y R_y) +$$

$$h.c.] + \sum_s \sum_k \{[\frac{t_2}{N} |k,D\rangle\langle k,A| + \frac{v}{N} \exp(ik_y(R_y + \Delta R_y))|k,A\rangle\langle k,D| \exp(-ik_y R_y) + h.c.]\}.$$

$$(3)$$

We make the replacement

$$\{|k,A\rangle, |k,B\rangle, |k,C\rangle, |k,D\rangle\} \rightarrow \{a_k, b_k, c_k, d_k\} \quad (4)$$

below, where $\{a_k, b_k, c_k, d_k\}$ the destruction operators in momentum space. We can then present the spinless, reciprocal space Hamiltonian in the basis $(a_k \; b_k \; c_k \; d_k)^T$ as

$$H_{NH,2D}(k_x, k_y) =$$

$$\begin{pmatrix} \varepsilon_1 & s & 0 & p \\ s^* & \varepsilon_2 & q & 0 \\ 0 & q^* & \varepsilon_1 & r \\ p^* & 0 & r^* & \varepsilon_2 \end{pmatrix}. \quad (5)$$

This reciprocal-space 2D SSH model is spinless. Therefore, the model is relevant to spin-polarized electrons and must be duplicated when applied to physical systems. Here, $\varepsilon_1 = i\gamma - \mu$, $\varepsilon_2 = -i\gamma - \mu$, $p = t_2 + v \exp(-iak_y)$, $s = u + t_1 \exp(iak_x)$, $q = v + t_2 \exp(iak_y)$,

$r = t_1 + u \, exp(iak_x)$, and $\mu$ is the chemical potential. The hopping parameters along the (horizontal) $x-$ and the (vertical) $y$ -directions are $(u, t_1)$ and $(v, t_2)$, respectively. The imaginary staggered potentials (ISP) on (A, B) and (C, D), respectively, are $(i\gamma, -i\gamma)$ and $(i\gamma, -i\gamma)$ are introduced additionally to make the Hamiltonian non-Hermitian albeit with balanced structure. The Hamiltonian $H_{NH,2D}(k_x, k_y)$ is $PT$-symmetric [6] if $s = r$, or $u = t_1$ [ $(PT) \, H_{NH,2D} \, (PT)^{-1} = H_{NH,2D}$] . Here, the inversion symmetry (IS) operator is $P = \sigma_x \otimes \sigma_x$ and the time-reversal symmetry (TRS) operator $T = I_4 \, K$ ; $I_4$ is 4 by 4 identity matrix and $K$ stands for complex conjugation. The inversion symmetry (IS)$(A \leftrightarrow D, B \leftrightarrow C)$ requires that $H_{NH,2D}(-k_x, -k_y) = P \, H_{NH,2D}(k_x, k_y) \, P^{-1}$ which is preserved if $u = t_1$, and $v = t_2$. We find that TRS is not preserved $[TH_{NH,2D}(k_x, k_y)T^{-1} \neq H_{NH,2D}(-k_x, -k_y)]$, as the ISPs $(\pm i\gamma)$ are involved in $H_{NH,2D}(k_x, k_y)$. In particular, for $u = t_1$, one may write (5) as $\begin{pmatrix} F & \beta \\ \beta^\dagger & F \end{pmatrix}$, where $F = \begin{pmatrix} \varepsilon_1 & s \\ s^* & \varepsilon_2 \end{pmatrix}$ and $\beta = \begin{pmatrix} 0 & p \\ q & 0 \end{pmatrix}$. As regards the chiral symmetry ( the simple guess for the corresponding operator is $\Gamma = \sigma_z \otimes \sigma_0, \Gamma^2 = I, \Gamma^{-1} = \Gamma$), we find that $H_{NH,2D}$ does not anti $-$ commute with $\Gamma$ unless $\gamma = 0$. We also find that if $\Gamma$ is assumed to be $\sigma_z \otimes \sigma_0$, $\{H_{NH,2D}, \Gamma\} \neq 0$, unless $r = -s$, and $p = -q^*$ which will trivialize the Hamiltonian. Thus, the Hamiltonian $H_{NH,2D}$ could not be made chiral symmetric. The particle-hole symmetry (PHS) $(C = \varsigma K, \varsigma = \sigma_x \otimes \sigma_z)$ is satisfied by $H_{NH,2D}(k_x, k_y)$ [ $\varsigma (H_{NH,2D}^T)^* \varsigma^{-1} = -H_{NH,2D}$], if $s = r$, and $q^* = p$. This implies that $H_{NH,2D}$ does not respect PHS when the horizontal bonds ($s$ and $r$) in Figure 1(a) do not match and the vertical bonds ($q$ and $p$) (in Figure 1(a)) also do not match under conjugation. Here, $\sigma_0$ and $\sigma_{x,y,z}$, respectively, are the two-by -two identity matrix and Pauli matrices. As we shall see in section 4 that the broken IS ($u \neq t_1$, and $v \neq t_2$) and the broken TRS ($\gamma \neq 0$) lead to non-zero Berry curvature and the anomalous Nernst effect.

**Energy eigenvalues**

The energy eigenvalues of (5) are given by $\det(H_{NH,2D} - \lambda I) = 0$. The determinant leads to the quartic $Q_{A,J,\gamma}(\lambda) = \lambda^4 - 2A(k_x, k_y)\lambda^2 + A^2(k_x, k_y) - J(k_x, k_y) = 0$, where $A = |p|^2 + |s|^2 - \gamma^2$, and $J = p^*(ps^* + qr)s + p(p^*s + q^*r^*)s^*$. The explicit expressions for the eigenvalues are $\lambda_1 = \sqrt{(A + \sqrt{J(k_x, k_y)})}$, $\lambda_2 = -\sqrt{(A + \sqrt{J(k_x, k_y)})}$, $\lambda_3 = \sqrt{(A - \sqrt{J(k_x, k_y)})}$ and $\lambda_4 = -\sqrt{(A - \sqrt{J(k_x, k_y)})}$. The left eigenvectors are

$$N_j^{-1/2} \times (s^*q^*r^* + p^*B_j \quad (\lambda_j - i\gamma)(p^*s + q^*r^*) \quad r^*C_j + p^*qs \quad (\lambda_j - i\gamma)(B_j - s^*s) \, ). \quad (6)$$

The right eigenvectors are

$$N_j^{-1/2} \times (sqr + pB_j \quad (\lambda_j - i\gamma)(ps^* + qr) \quad rC_j + pq^*s^* \quad (\lambda_j - i\gamma)(B_j - s^*s) \, )^T. \quad (7)$$

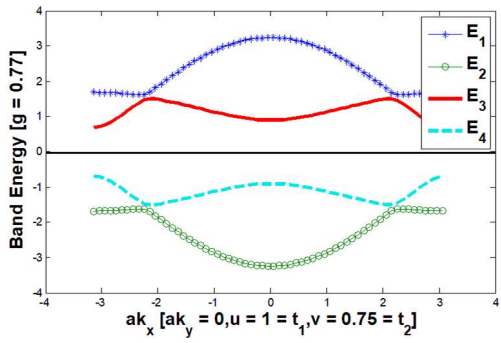

(a)

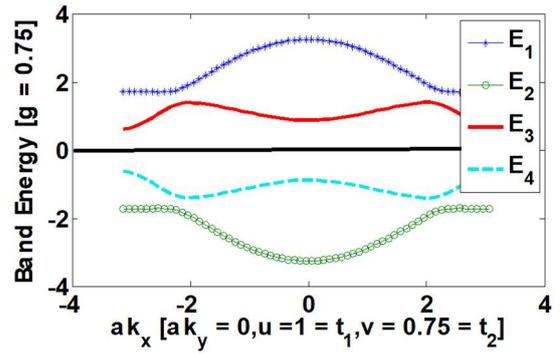

(b)

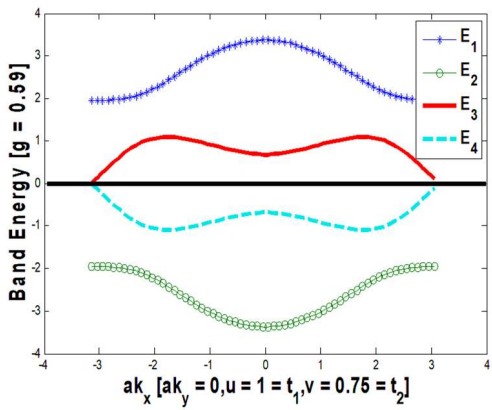

(c)

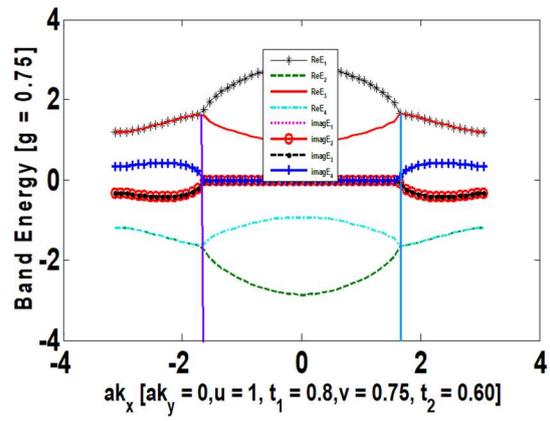

(d)

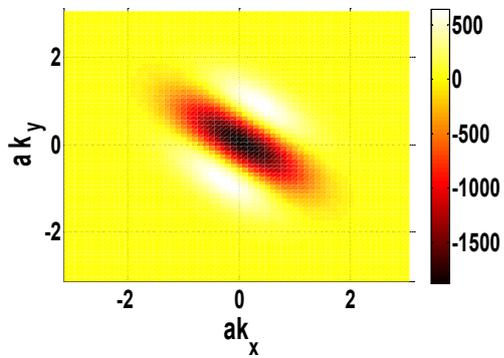

(e)

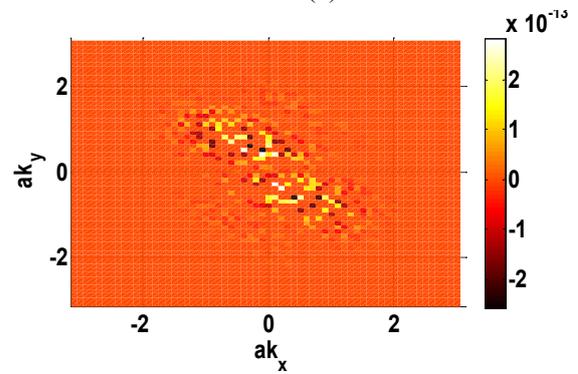

(f)

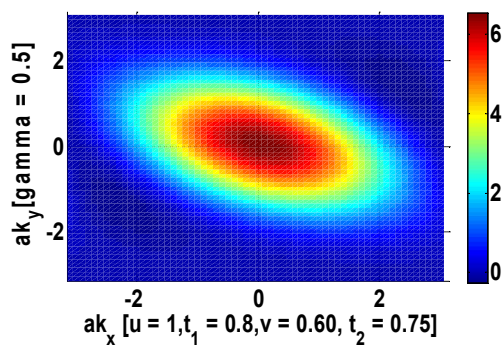

(g)

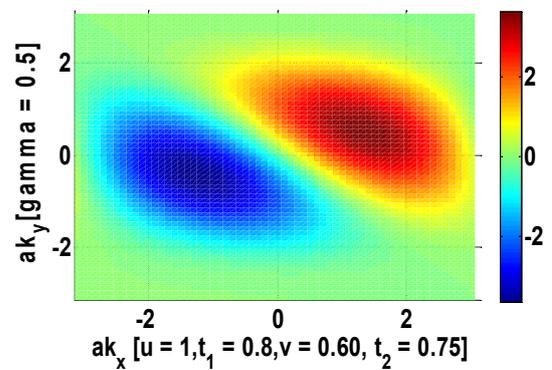

(h)

**Figure 2.** The plots of the energy eigenvalues $\lambda_j$ ($j = 1,2,3,4$), denoted as $E_j$, as a function of $ak_x$ with $ak_y = 0$. The parameter values in **(a) ((b))** are $u = t_1 = 1$, $v = t_2 = 0.75$, $\mu = 0$, and $\gamma = 0.77(0.75)$. In Figure **(c)**, however, $\gamma = 0.59$. Here the symbol 'g' is the shortform of 'gamma($\gamma$)' in our 2DSSH model. The horizontal solid line represents the Fermi energy. **(d)** The plots of the real and the imaginary parts of the energy eigenvalues as a function of $ak_x$ with $ak_y = 0$. The parameter values used in Figures **2(d)** are $u = 1, t_1 = 0.80$, $v = 0.75, t_2 = 0.60$, $\mu = 0$, and $\gamma = 0.75$. The solid horizontal lines correspond to $\mu = 0$. **(e)((f))** The plot of the real (imaginary) part of the discriminant $\Delta(k_x, k_y)$ of the quartic $Q(\lambda) = 0$. **(g) ((h))** The plot of the real (imaginary) part of one of the $3 \times 3$ minors $[p^*(ps^* + qr)]$ as a function of $ak_x$ and $ak_y$.

Here $B_j = \lambda_j^2 + \gamma^2 - q^*q$, and $C_j = \lambda_j^2 + \gamma^2 - s^*s$. Upon using the bi-orthogonality condition, we obtain $N_j$. Figure 2 presents energy eigenvalues plotted as a function of $ak_x$ with $ak_y = 0$. In Figure 2 (a) (2(b)), the parameter values are $u = t_1 = 1$, $v = t_2 = 0.75$, $\mu = 0$, and $\gamma = 0.77(0.75)$. However, in Figure (c), $\gamma$ equals 0.59. We have chosen '$u$' as the unit of energy throughout this paper. Our graphical representations in Figures 2(a) – 2(c) demonstrate that the tunable gain/loss parameter $\gamma$ transforms the system from an insulator to a conductor when its numerical value is decreased. An important observation is the presence of gapless and gapped approximate Dirac cones in Figures 2(a) and (b)[21]. $H_{NH,2D}$ is PT-symmetric (eigenvalues are real) if $u = t_1$ in a limited region of two-dimensional BZ. In contrast, for $u \neq t_1$, the Hamiltonian yields complex eigenvalues, indicating a transition to a broken PT-symmetric phase. Figure 2(d) plots the real and imaginary parts of energy eigenvalues as a function of $ak_x$ with $ak_y = 0$, where $u \neq t_1$ (and $v \neq t_2$). The parameter values used are $u = 1, t_1 = 0.8$, $v = 0.75, t_2 = 0.60$, $\mu = 0$, and $\gamma = 0.75$. The choice corresponds to broken PT and PH symmetry. The figure indicates that at momenta close to $\pm 2$, two or more eigenvalues (both real and imaginary parts) and their corresponding eigenvectors of the system converge. The Figure 2(d) indicates that coalescence may occur for certain momentum values, as shown by vertical lines. The solid horizontal lines correspond to $\mu = 0$. Furthermore, one of the conditions for exceptional points (EPs) is that the discriminant of the quartic in $Q_{A,J,\gamma}(\lambda)$ above must be zero. This condition picks out coalescing eigenvalues — a necessary condition for EPs. For the quartic $Q_{A,J,\gamma}(\lambda) = 0$ with complex coefficients, one can (still) write its discriminant $\Delta(k_x, k_y) = 0$ for equal roots. In Figures 2(e), and 2(f), we have contour plotted the real and imaginary parts of $\Delta(k_x, k_y)$ as a function of $(ak_x, ak_y)$. The plots indicate that $\Delta(k_x, k_y) \to 0$ at large number of points in the $ak_x - ak_y$ plane.

We are now interested in exceptional points (EPs) of $H_{NH,2D}$. We notice that when $J(k_x, k_y) = 0$, $Q_{A,J,\gamma}(\lambda) = Q_{A,\gamma}(\lambda) = (\lambda^2 - A(k_x, k_y))^2 = 0$ implies a square-root branch point structure, typical of multi-level EP physics. There are double roots at $\lambda_\pm(k_x, k_y) = \pm\sqrt{A(k_x, k_y)}$ with algebraic multiplicity (AM) as 2. In view of Eqs. (6) and (7), the eigenvector coalescence occurs in this case. The geometric multiplicity (GM) of the eigenvalues, on the other hand, is the dimension of the eigenspace corresponding to these eigenvalues, i.e. the number of linearly independent eigenvectors associated with $\lambda_\pm(k_x, k_y)$. Mathematically, GM $(\lambda_\pm) = \dim(\ker(H_{SSH,2D} - \lambda_\pm I))$. This is the nullity of $(H_{SSH,2D} - \lambda_\pm I)$. By the rank-nullity theorem, nullity(matrix) = rank(matrix)+number of columns of the matrix. So, in this case, nullity = $[4 - rank(H_{SSH,2D} - \lambda_\pm I)] <$ AM of $\lambda_\pm(k_x, k_y)$ is describing the case when GM < AM. To find the rank, we need to compute the determinants of all minors of $(H_{SSH,2D} - \lambda I)$, starting from the 4×4 determinant down to 1×1. If **det** $(H_{SSH,2D} - \lambda I) \neq 0$, then rank = 4. In this case, $\lambda$ is not an eigenvalue of $H_{SSH,2D}$. However, if **det** = 0, we need to check all 3×3 minors. If one is nonzero, rank = 3. Upon going down, if all 3×3 minors vanish, one needs to check 2×2 minors. If all 2×2 minors vanish, one needs to check 1×1 entries. The largest size minor with nonzero determinant gives the rank. In Figures 2(g) and 2(h), we have plotted the real and imaginary parts of one of the 3×3 minors [ $p^*(ps^* + qr)$]. The plots show the possibility of the minor being non-zero at large number of points in the $ak_x - ak_y$ plane. This implies that GM = 1, or GM < AM. Thus, the Hamiltonian $H_{SSH,2D}$ becomes defective (non-diagonalizable), leading to potential EPs (with AM = 2) at $\lambda_\pm(k_x, k_y) = \pm\sqrt{A(k_x, k_y)}$.

Next, we analyze the phase rigidity factor (PRF) $P_j = |\langle v^{(j)}|u^{(j)}\rangle|/|\langle u^{(j)}|u^{(j)}\rangle|$ graphically to identify some of the true EPs in the $ak_x - ak_y$ plane. In Figures 3(a) - 3(c), we have plotted $P_1(\lambda_+)$ and $P_2(\lambda_-)$ as functions of $\gamma$ for $\mu = 0$ involving staggered hopping amplitudes (u $\neq t_1$ and v $\neq t_2$).The parameter values used are $u = 1, t_1 = 0.80, v = 0.60,$ and $t_2 = 0.71$. The choice of the parameter values breaks PT- and PH-symmetry. In Figures 3(d) - 3(f), we have plotted the same with the parameter values $u = t_1 = 1$, and $v = t_2 = 0.75$. The choice of the values leaves PT- and PH-symmetry unbroken. As we find in these figures, there are quite a few points $(ak_x, ak_y)$ where $P_1(\lambda_+)$, and/or $P_2(\lambda_-) \to 0$, for example $(ak_x = \pi, ak_y = -\pi)$ for $-0.3 < \gamma < +0.3$ as in Figure 3(a), $(ak_x = \pi, ak_y = -\pi/2)$ for $\gamma \sim \pm 1$ as in Figure 3(b), and $(ak_x = \pi, ak_y = 0)$ for $\gamma \sim \pm 1.5$ as in Figure 3(c). In the PT- and PH-symmetry unbroken case, as exemplified by Figures 3(d), 3(e), and 3(f), $P_1(\lambda_+)$, and/or $P_2(\lambda_-) \to 0$ for $\gamma \sim 0, \gamma \sim \pm 1$, and $\gamma \sim \pm 1.5$, respectively. We, thus, find that the use of the rank-nullity theorem and the graphical analysis of the phase rigidity factor permits the identific-

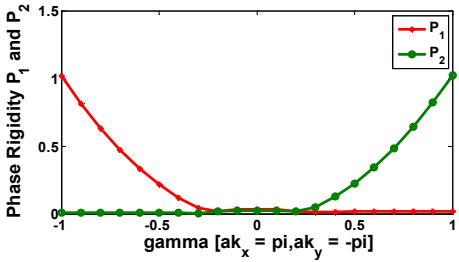

(a)

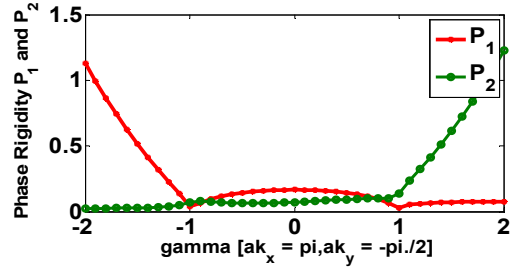

(b)

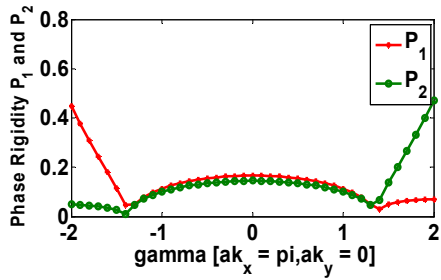

(c)

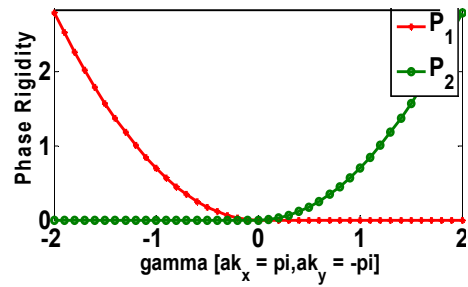

(d)

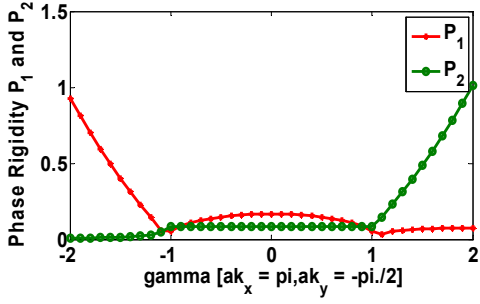 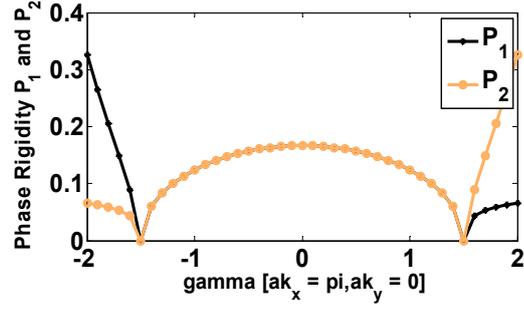

(e)　　　　　　　　　　　　　　　(f)

**Figure 3.** (a),(b),(c) The plots of $P_1(\lambda_+)$ and $P_2(\lambda_-)$ as functions of $\gamma$ for $\mu = 0$ involving staggered hopping amplitudes (u $\neq t_1$ and v $\neq t_2$). The parameter values used are $u = 1, t_1 = 0.80,\ v = 0.60,$ and $t_2 = 0.71$. In Figures 3(d) - 3(f), there are plots of the same with the parameter values $u = t_1 = 1,$ and $v = t_2 = 0.75$. The choice of the values leaves PT- and PH-symmetry unbroken.

ation of such points. Unique to non-Hermitian systems, these points result in enhanced sensitivity to tiny changes in system parameters (e.g. gain/loss, and hopping terms) and random perturbations (e.g. weak disorder in hopping amplitudes, and alteration in boundary conditions), triggering dramatically amplified responses in the system's behavior.

## 3. Zak Phase and Topolectric RLC Circuit

### (A) Zak Phase

For non-Hermitian systems, the Zak phase undergoes modification due to the loss of time-reversal symmetry and the possibility of complex eigenvalues. In particular, the Zak phase for a 2D non-Hermitian SSH model represents a generalization of the 1D case (the Zak phase in 1 D is essentially an integral, defined as $\phi = \oint dk\, A(k)$ where $A(k)$ is the Berry connection), incorporating complex hopping terms and non-Hermitian effects like dissipation. The Zak phase in this scenario is a vector-like quantity, consisting of two components, one for each direction in the 2D system. Furthermore, non-Hermitian systems with a nonzero Zak phase may exhibit the 'non-Hermitian skin effect' **[24–36]**, where bulk states localize at the system's boundaries due to non-Hermitian effects. In this communication, by utilizing the Wilson loop approach **[37,38]** and combining methodologies from references **[39,40]**, we have calculated the vectorized Zak phase below.

As noted above, Berry connections are essential in the context of the Zak phase. In a multi-band system, the total Berry connection cannot be expressed as a simple sum of individual Berry connections for each band, as it is contingent upon the specific wavefunctions of the bands and the inter-band couplings, which cannot be easily separated. However, if the bands are energetically well-separated and exhibit minimal inter-band coupling, the Berry connection for each band may be treated as approximately independent. We employ the parameter values close to those in Figures 2 to calculate the Zak phase below for $\gamma = 0$, and $\gamma = 0.50$. We first consider the function $\phi_x(k_y)$ for $\gamma = 0$. In this case, the Berry connection at $k_x = k_\alpha$ (where we have assumed $k_\alpha = \pi$ in our calculation and $\alpha = 1,\ldots,N \gg 1$) and for the $j^{th}$ band is given by $A^{(j)}(k_\alpha, k_y) = \langle u^{(j)}(k_\alpha, k_y)|i\partial_k|u^{(j)}(k_\alpha, k_y)\rangle$. Here $|u^{(j)}(k_\alpha, k_y)\rangle$ is the Bloch function and $j$ now runs over all the bands below the band gap. The Zak phase for the $j^{th}$ Bloch band $\phi_\alpha^{(j)}(k_y)$ in a small segment connecting $k_\alpha$ and $k_{\alpha+1}$ is $\phi_\alpha^{(j)}(k_y) = A^{(j)}(k_\alpha, k_y)\Delta(ak_\alpha) = i\langle u^{(j)}(k_\alpha, k_y)|u^{(j)}(k_{\alpha+1}, k_y)\rangle - i\langle u^{(j)}(k_\alpha, k_y)|u^{(j)}(k_\alpha, k_y)\rangle$. In our numerical calculation, we have assumed $\Delta(ak_\alpha) = \Delta(ak) \ll 1$. We have used Matlab package for the numerical calculation. It follows that $\langle u^{(j)}(k_\alpha, k_y)|u^{(j)}(k_{\alpha+1}, k_y)\rangle = 1 -$

$i\phi_\alpha^{(j)}(k_y) \approx exp(-i\phi_\alpha^{(j)}(k_y))$. In view of the Wilson loop approach[37,38], the total Zak phase $\phi_x(k_y)$ could be calculated by compounding the discrete Zak phase from each small segment $\Delta(ak_\alpha)$. We find

$$exp(-i\phi_x(k_y)) = \prod_{\alpha=1}^{\alpha=N} \Sigma_j \langle u^{(j)}(k_\alpha, k_y)|u^{(j)}(k_{\alpha+1}, k_y)\rangle \quad (11a)$$

following refs. [66]. Upon using this expression, the Zak phase component $\phi_x$ could be written as

$$\phi_x = (\frac{1}{2\pi}) \int_{-\pi}^{+\pi} Im[\ln\{\prod_{\alpha=1}^{\alpha=N} \Sigma_j \langle u^{(j)}(k_\alpha, k_y)|u^{(j)}(k_{\alpha+1}, k_y)\rangle\}]dk_y \quad (11b)$$

The other component of the vectorized Zak phase, viz. $\phi_y$, can be obtained in a similar manner. In non-Hermitian 2D systems ($\gamma \neq 0$) also, we have the two components of the Zak phase, viz. $\{\phi_x(k_y), \phi_y(k_x)\}$ defined with the help of right and the left eigenvectors ($|u^{(j)}\rangle, \langle v^{(j)}|$). We have plotted the vectorized Zak phase components $\phi_x$ and $\phi_y$ as a function of $\frac{u}{v}$ in Figures 4(a) − 4(d). The plots indicate that the conventional bulk-boundary correspondence depends crucially on the ratio $\frac{u}{v}$. In Figures 4(a) and 4(b), we have assumed $\gamma = 0$, whereas in 4(c) and 4(d) we have assumed $\gamma = 0.50$. In these figures, the hopping parameters $u = t_1$ and $v = t_2$. Furthermore, we note that, here $\phi_x$ and $\phi_y$ approximate $\pi = 4\arctan(1) \approx 3.1416$ (correct to 4 decimal places) with relative error $\eta = 0.0021$. The plots reveal that the conventional bulk-boundary correspondence hinges significantly on the ratio $u/v$.

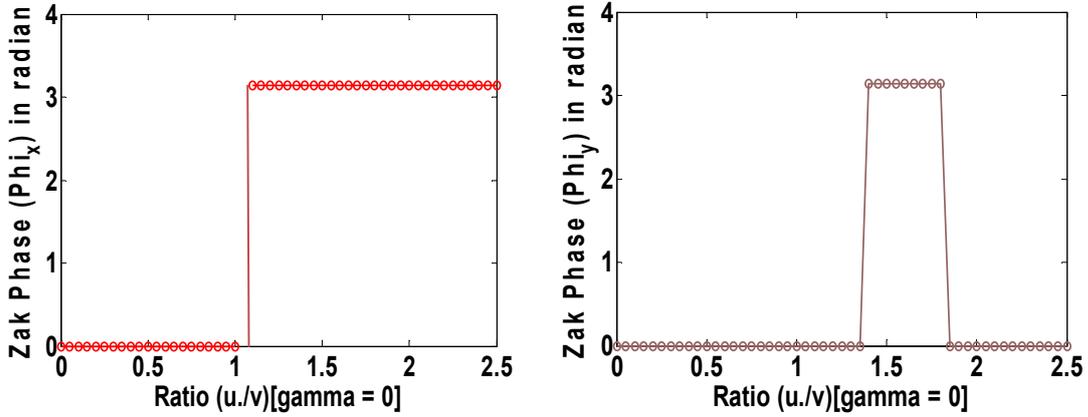

(a)                                  (b)

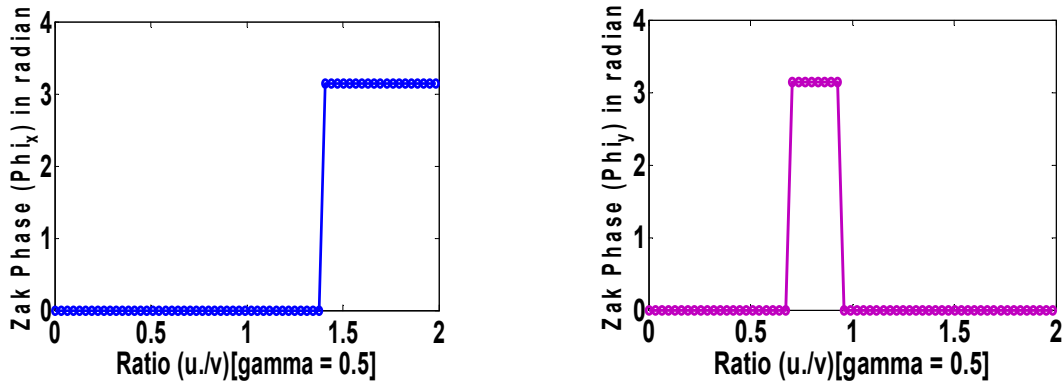

**(c)**  **(d)**

**Figure 4.** The plots of the vectored Zak phase components $\phi_x$ and $\phi_y$ as a function of $\frac{u}{v}$. In these figures, the hopping parameters $u = t_1$ and $v = t_2$ ensuring inversion symmetry protection. In Figures 4(a) and 4(b), we have assumed $\gamma = 0$, whereas in 4(c) and 4(d) we have assumed $\gamma = 0.50$.

## (B) Topolectric RLC Circuit

An RLC circuit arranged in a lattice-like structure, with inductors, capacitors, and resistors forming a periodic configuration as shown in Figure 1(b), is under consideration. The relationships between voltage and current in this circuit are governed by impedance, which is a function of frequency, and the energy stored in the inductors and capacitors. The equations of motion for the voltages across the capacitors and the currents through the inductors can be written in a matrix form, where the circuit Laplacian is related to the impedance of the circuit. The circuit Laplacian is utilized to describe the dynamics of the voltages in the system, typically in the form of a differential equation that governs the time evolution of the voltages across the capacitors and the currents through the inductors. The Laplacian in a topolectric circuit can be viewed as analogous to the Hamiltonian in condensed matter systems, necessitating the solution for eigenfrequencies and eigenstates of this matrix, where eigenvalues of the matrix correspond to oscillation frequencies and eigenstates describe voltage or current distributions across components. This exercise will also enable one to calculate the Zak phase, say for the topolectric circuit in 1(b), to identify topological characteristics, such as protected edge states, akin to condensed matter physics. The calculation is relegated to a future communication.

The preliminary investigation of the 2D topolectric circuit, presented in this section, was motivated by refs. **[40-44]**. We, however, begin with a simple LC system, featuring unit cells consisting of a pair of capacitors $C_1$ and $C_2$ with identical inductors L between them and an alternating driving voltage source with frequency $\omega$, subject to periodic boundary conditions, characterized by the grounded Laplacian of the form $J_{SSH}(k) = i\omega \left( C_1 + C_2 - \frac{1}{\omega^2 L} \right) \sigma_0 - i\omega \left[ ( C_1 + C_2 \cos(ak)) \sigma_x + (C_2 \sin(ak))\sigma_y \right]$. The $\sigma_j$'s are Pauli matrices and $(ak)$ is the inverse modulation wavelength. The Laplacian has a strong resemblance with the 1D-SSH model. The topological boundary resonance (TBR) condition, which essentially corresponds to the secular equation of the system, yields $\omega = \sqrt{\frac{1}{LC_{eff}}}$ where $C_{eff} = C_1 + C_2 \pm \sqrt{C_1^2 + C_2^2 + 2C_1 C_2 \cos(ak)}$. In what follows we utilize the same approach to analyze the system shown in 1(b).

The mapping of the current flow in a topoelectric circuits with a network of elements like resistors, inductors and capacitors, in a form similar to a tight-binding Hamiltonians of 1D/2D SSH model, is possible **[40-44]** where resistors (R), capacitances (C) and inductances (L) act as the hopping parameters. We consider below a topolectric $RLC$ circuit as shown in Figure 1. The resistors $R$ (not shown) are assumed to be in parallel with the capacitors (C). This figure depicts nodal points with the voltages $V_1, V_2, V_3,$ and $V_4$ corresponding to the four nodes 1,2,3, and 4, respectively. Upon using Kirchoff's laws, the current flowing out of a node $j = 1,2,3,4$ could be written as

$$I_{N,j} = \sum_{k \neq j} Y_L \left( V_j - V_k \right) + Y_C V_j \qquad (12a)$$

where $\sum_{k \neq j} Y_L \left( V_j - V_k \right)$ is equal to the current flowing out of the node $j$ to other nodes $k$ inked by admittance $Y_L = -\frac{i}{\omega L}$ and the current flowing to ground with admittance $Y_C = i\omega C + R^{-1}$. The driving voltage of the circuit may be assumed to have frequency ω. Upon using the formulation presented in references **[40,43]**, we establish below link of with topological systems: In the case of lattice systems, the indexing of eigenmodes by momentum $\mathbf{k}$ and band index $\alpha$ is possible using Bloch's theorem. Similarly, we write here for the eigenmode $\xi_{k,\alpha}(\mathbf{r}, \beta) = \varsigma_\alpha(\mathbf{k}, \beta) \exp(i\mathbf{k} \cdot \mathbf{r})$, where a node $(N, j)$ (may be referred to as a lattice site analogous to a lattice system) is indexed by its position $\mathbf{r}$ and sublattice (if any) label $\beta$. The admittance between two sites $(\mathbf{0}, \beta)$ and $(\mathbf{r}, \beta')$ takes the form

$$Y_r^{\beta\beta'} = \{\sum_{k,\alpha} \frac{|\varsigma_\alpha(\mathbf{k},\beta) - \varsigma_\alpha(\mathbf{k},\beta')\exp(i\mathbf{k}\cdot\mathbf{r})|^2}{J_{k,\alpha}}\}^{-1}. \qquad (12b)$$

The inverse wavelengths $\mathbf{k} = (k_x, k_y)$ take care of spatial modulation to the circuit elements in the two directions of the circuit and the $J_{k,\alpha}$ is circuit Laplacian matrix ($J$) element to be introduced below for a RLC circuit. In fact, for such circuits, the voltage ($V$) and current ($I$) column vectors are related as $J\,V = \lambda\,V = I$. Upon comparing this fundamental circuit equations with the Schrodinger Equation $H\,\psi = E\,\psi$, both in explicit matrix form, we notice that $J$ has a role similar to the Hamiltonian $H$. Thus, if the (dimensionless) admittance between two nodes $(\mathbf{0}, \beta)$ and $(\mathbf{r}, \beta')$ becomes much less than one, there exist nontrivial eigenstates with the eigenvalues $\lambda \to 0$. This drastic increase of the impedance corresponds to the robust topological boundary resonance (TBR) condition in RLC circuits. The noting above provides us with a valuable insight that a formulation starting from circuit theory may allow us to have a sneak peek of unusual quantum phenomena.

One can write the current flowing out of the four nodes in Figure 1(b) in the explicit matrix form as $I = i\omega J V$ with $J = \begin{pmatrix} F_1 & \varsigma_1 \\ \varsigma_1^\dagger & F_1 \end{pmatrix}$, where

$$F_1 = \begin{pmatrix} b & a_2 + a_1 \exp(-ik_x) \\ a_2 + a_1 \exp(ik_x) & b \end{pmatrix}, \varsigma_1 = \begin{pmatrix} 0 & a_2 + a_1 \exp(-ik_y) \\ a_2 + a_1 \exp(-ik_y) & 0 \end{pmatrix}.$$

$$(13)$$

The column matrices $I$ and $V$ are given below:

$$I = \begin{pmatrix} I_{N,1}(k_x, k_y) \\ I_{N,2}(k_x, k_y) \\ I_{N,3}(k_x, k_y) \\ I_{N,4}(k_x, k_y) \end{pmatrix}, V = \begin{pmatrix} V_1 \\ V_2 \\ V_3 \\ V_4 \end{pmatrix}, a_1 = \frac{1}{\omega^2 L_1}, a_2 = \frac{1}{\omega^2 L_2}, b = \left(C - \frac{i}{R\omega} - 2a_1 - 2a_2\right). \quad (14)$$

A comparison of Eq. (13) with Eq. (5) shows that the former has some degree of similarity with the 2D SSH model except that a non-zero R leads to the dissipation only. The eigenvalues $\lambda$ of the circuit Laplacian matrix given above in (13) are obtained quite easily. The imaginary part of the secular equation yields $C - 2a_1 - 2a_2 = 0$. This gives $\omega = \sqrt{\frac{2}{(L_1+L_2)C}}$. For the parameter values $L_1 = 10\,mH = L_2$, and $C = 0.01\mu F$, we obtain $\omega = 10^5\,s^{-1}$. The real part of equation, however, involves the inverse modulation wavelengths $(k_x, k_y)$. The real part yields

$$E(k_x, k_y) = C - 2a_1 - 2a_2 \pm \sqrt{(\tfrac{1}{(R\omega)^2} + P \pm \sqrt{P^2(k_x,k_y) - Q^2(k_x,k_y)}\,)}, \qquad (15)$$

where

$$P(k_x, k_y) = 2a_1^2 + 2a_1 a_2 (\cos(k_x) + \cos(k_y)) + 2a_2^2,$$

$$Q^2(k_x, k_y) = \sum_{j=(x,y)} (2a_1^2 + 2a_1 a_2 \cos(k_j) + 2a_2^2)^2 - (a_1^2 + 2a_1 a_2 \cos(k_y) + a_2^2) \times$$
$$(2a_2^2 + 4a_1 a_2 \cos(k_x) + 2 a_1^2 \cos(2k_x)). \quad (16)$$

The plots of $E(k_x, k_y)$ as a function of $\omega$, for the parameter values $L_1 = 0.1 \, mH = L_2$, and $C = 0.01 \mu F$, are given below in Figure 5 for R = (1, 26, 50) Ohm. These plots show that the topological boundary resonance (TBR) condition $E = 0$ in RLC circuit is satisfied when R ~ 1 Ohm or less. The boundary resonance occurs at $\omega \sim 3 \times 10^4 \, s^{-1}$ for all four eigenvalues obtained equating Real($E$) with zero. The value of the inverse wavelengths $\boldsymbol{k} = (k_x, k_y)$, which take care of spatial modulation to the circuit elements in the two directions of the circuit, have no significant influence on the value of $\omega$. For higher resistances (as shown in Figures 5(c)-5(f)), $(E_1, E_2)$ do not become zero. One may draw the conclusion that the compliance of the robust TBR condition is possible in a RLC topoelectric circuit when $R \leq 1 \, Ohm$. In other words, a larger $R$ does not favor TBR. As noted in ref. **[40],** the topological boundary resonances remain robust even in the face of significant nonuniformity of circuit elements. This promises high-precision identification in a realistic measurement. With series resistance $R$ on the inductor, one needs to replace the impedance of each inductor replaced by $i\omega L \rightarrow i\omega L + R$.

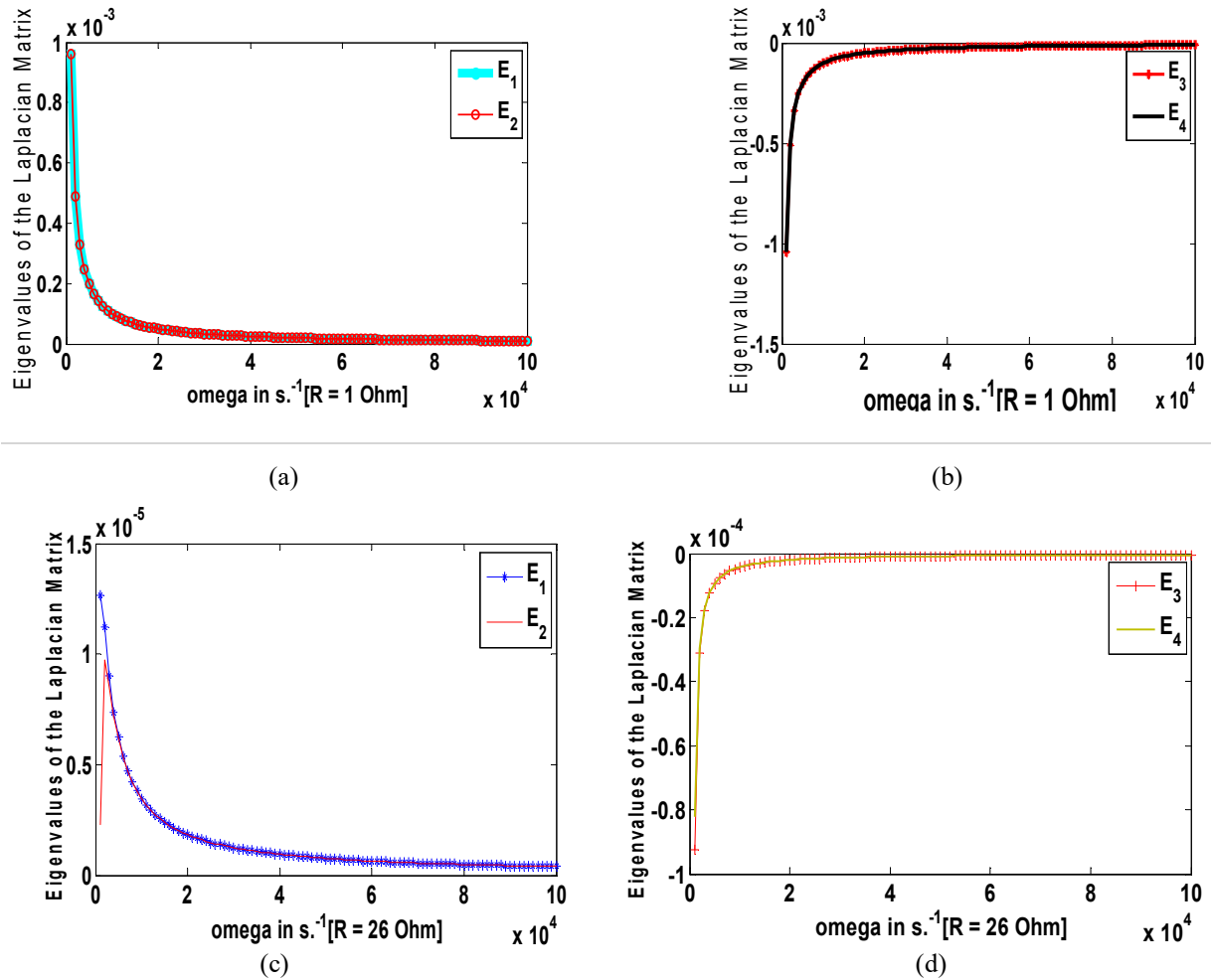

(a)

(b)

(c)

(d)

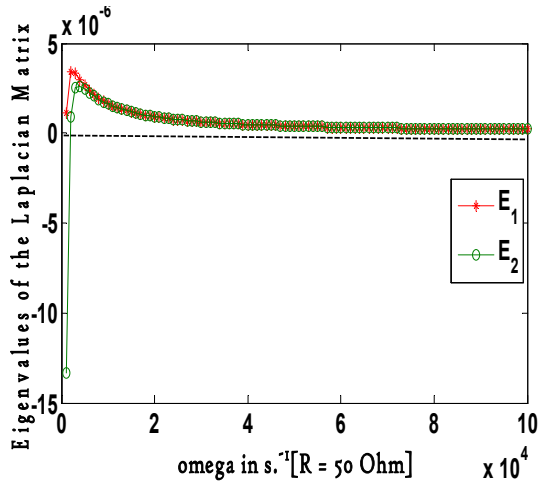
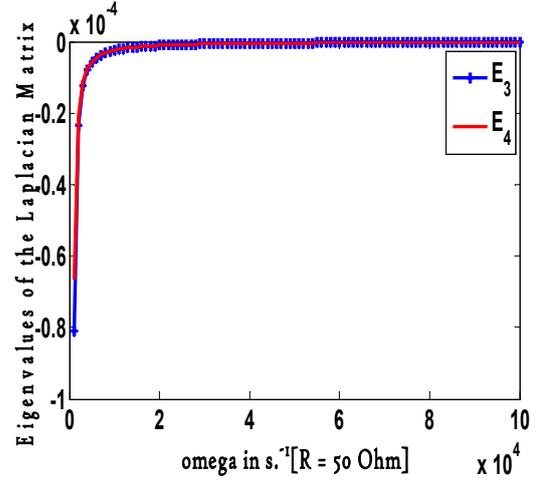

(e)                    (f)

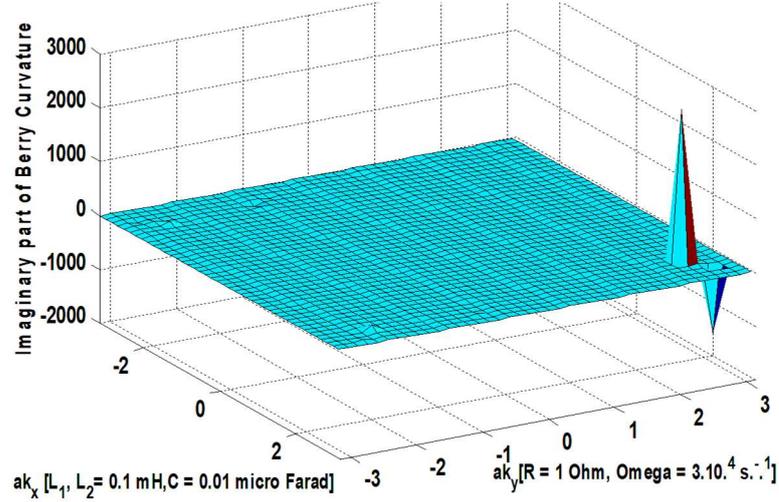

(g)

**Figure 5. (a)-(f)** The plots of the real part of eigenvalues of the circuit Laplacian matrix in (13) as a function of $\omega$ in $s^{-1}$. The parameter values are $L_1 = 0.1\ mH = L_2$, $C = 0.01 \mu F$, and R = (1, 26,50) Ohm. **(g)** A plot of the imaginary component of the Berry phase for $L_1 = 0.1\ mH = L_2$, $C = 0.01 \mu F$, $R = 1$ Ohm, and $\omega \sim 3 \times 10^4\ s^{-1}$. The component comes from the non-Hermitian part of the circuit Laplacian.

In Figure 5(g), we have also presented a plot of the imaginary component of the Berry phase [see ref.**(45)**] as a function of the inverse modulation wavelengths $\mathbf{k} = (k_x, k_y)$. This originates from the non-Hermitian (dissipative) part ($\frac{i}{R\omega}$) of the circuit Laplacian. The parameter values used in the plot are $L_1 = 0.1\ mH = L_2$, $C = 0.01 \mu F$, $R = 1$ Ohm, and $\omega \sim 3 \times 10^4\ s^{-1}$. Upon integrating ICBP numerically ($-\pi \leq (k_x, k_y) \leq \pi$), we find that the quantum Hall susceptance $\Im \sim 0.92$ is positive which indicates that the system has capacitive properties. It is possible to have negative $\Im$, by tuning R, in which case the system has inductive properties. The $\mathbf{k}$ − integration process is facilitated utilizing the Matlab package, wherein the complexity of the integration is mitigated by dividing k-space into smaller grids, enabling the integration to be performed in a manageable manner, and our approximations are refined as additional grids are incorporated. Our study in this section has successfully implemented vectorized Zak phase quantization and analyzed the topolectric RLC circuit in Figure 1(b) to determine the corresponding topological boundary resonance condition and the quantum Hall susceptance ($\Im$).

## 4. Staggered Hopping Amplitudes

he presence of non-zero Berry curvature in systems with broken TRS/IS gives rise to an anomalous velocity, generally, resulting in an anomalous transport current and intrinsic Hall conductivity. In this section, apart from disrupted TRS ($\gamma \neq 0$), we investigate a scenario where the hopping amplitudes are staggered ($u \neq t_1$, and/or $v \neq t_2$) in the horizontal and vertical($\hat{x}$ and $\hat{y}$) – directions in Figure 1(a) causing broken IS and PHS with $\gamma = 0$. It must be emphasized that, apart from broken symmetries, an essential requirement for obtaining the integer Chern number is the existence of a band-gap at $\Gamma$ (0,0)between the bands closest to the Fermi energy. If this gap does not exist, the calculation of the Chern number becomes meaningless. The PT symmetric situation, depicted in Figures 2(a) and 2(b) (with $\gamma = 0.77(0.75)$), satisfies this necessary requirement under certain parameter values. In what follows we show that BC is finite leading to the anomalous Nernst response (ANE) in the Hermitian system ($\gamma = 0$) with, where the hopping amplitudes are staggered, and the broken TRS system ($\gamma \neq 0$). The finite BC emerges despite the fact that there is no spin-orbit interaction (SOI) to connect the two copies of the present SSH model describing spin-polarized electrons.

The quantum geometric tensor (QGT)**[46-51]** $G_{\mu\nu}$ is a matrix that captures the geometry of the quantum wavefunctions in parameter space (here, momentum space $\mathbf{k}$) and is defined as $G_{n,\mu\nu}(\mathbf{k},\lambda) = \left\langle \frac{\partial u_{n,k}(\lambda)}{\partial k_\mu} \middle| \frac{\partial u_{n,k}(\lambda)}{\partial k_\nu} \right\rangle - \left\langle \frac{\partial u_{n,k}(\lambda)}{\partial k_\mu} \middle| u_{n,k}(\lambda) \right\rangle \left\langle u_{n,k}(\lambda) \middle| \frac{\partial u_{n,k}(\lambda)}{\partial k_\nu} \right\rangle$ for a Hermitian system where the symbol $|u_{n,k}(\lambda)\rangle$, which smoothly depends on the *N*-dimensional parameter $\lambda = (\lambda_1, ..., \lambda_N)$, stands for the nth eigenenergy of a quantum Hamiltonian *H* and for a given momentum $\mathbf{k}$. The real part of the QGT is the quantum metric tensor (QMT) $g_{n,\mu\nu}(\mathbf{k},\lambda)$ which defines the distance between two quantum states. Its imaginary part corresponds to the Berry curvature $\Omega_{n,\mu\nu}(\mathbf{k},\lambda) = -2 \operatorname{Im} G_{n,\mu\nu}(\mathbf{k},\lambda)$. It is related to the topological properties of the system. The BC and QMT can also be written in terms of the derivatives of the Hamiltonian **[46-51]**. In a non-Hermitian system, QGT may still be defined but one has to make the left eigenstates orthonormal to the right eigenstates $\langle u^L(\lambda)|u^R(\lambda)\rangle = \delta_{R,L}$. This biorthonormality condition is a choice of normalization and not automatic. For such systems, the BC and QMT must be modified by considering the left and the right eigenstates. Here, we calculate below the z-component of BC $\Omega_\alpha^{\mu,\nu}(k_x, k_y)$ for the α$^{th}$ occupied band given by the formula

$\Omega_\alpha^{\mu,\nu}(k_x, k_y) =$
$\left[ i \sum_{\beta \neq \alpha} (\lambda_\alpha(k_x, k_y) - \lambda_\beta(k_x, k_y))^{-2} \left\{ \left\langle u^{(\alpha,\mu)}(k_x, k_y) \middle| \frac{\partial H_{NH,2D}(k_x,k_y)}{\partial k_x} \middle| u^{(\beta,\nu)}(k_x, k_y) \right\rangle \right.\right.$
$\left.\left. \times \left\langle u^{(\beta,\mu)}(k_x, k_y) \middle| \frac{\partial H_{NH,2D}(k_x,k_y)}{\partial k_y} \middle| u^{(\alpha,\nu)}(k_x, k_y) \right\rangle - (x \leftrightarrow y) \right\} \right]$,

(17)

where the energy eigenvalues $\lambda_\alpha(k_x, k_y)$ and $\mu, \nu = L, R$ represent the left and right eigenstates, subject to the normalization condition mentioned in Section 2. Notably, the four Berry curvatures - left-right (LR), right-left (RL), left-left (LL), and right-right (RR) - each providing a unique definition of BC exhibit local differences, but ultimately yield the same Chern number upon integration [52]. Equation (17) is what one obtains from QGT and, for a Hermitian system, it is the same as the conventional Kubo formula [53,54]. The outline of the derivative calculation is given in Appendix A. In Figure 6(a) and 6(b), we have shown the plots of BC in the z-direction for the staggered hopping amplitudes. The numerical values of the parameters used in the plots are $u = 1, t_1 = 0.95, t_2 = 0.5$, $v = 0.23 \,(0.35), \mu = 0$, and $\gamma = 0$. With these values, though TRS is respected, the inversion symmetry, however, gets disrupted. The band-gap, between the bands closer to the Fermi energy used in the plots, is $2\sqrt{(A - \sqrt{J(k_x, k_y)})}$. On the other hand, the magnitude of the gap between the occupied bands is $\{\sqrt{(A + \sqrt{J(k_x, k_y)})} + \sqrt{(A - \sqrt{J(k_x, k_y)})}\}$. In order to investigate Berry-phase effect in anomalous transport, we will first consider the anomalous Nernst conductivity (ANC) $\alpha_{xy}(\mu, T)$. The conductivity can be computed by integrating Berry curvature with entropy density over first BZ **[55,56]**. In the low-temperature limit, upon using the Mott relation **[55]** we obtain

$$\alpha_{xy} \approx \frac{\frac{\pi^2}{3} k_B^2 T}{e} \sum_n \int d\mathbf{k}\, \Omega_{xy}^{(n)} \frac{\partial f(E_n(k))}{\partial E_n(k)}. \qquad (18)$$

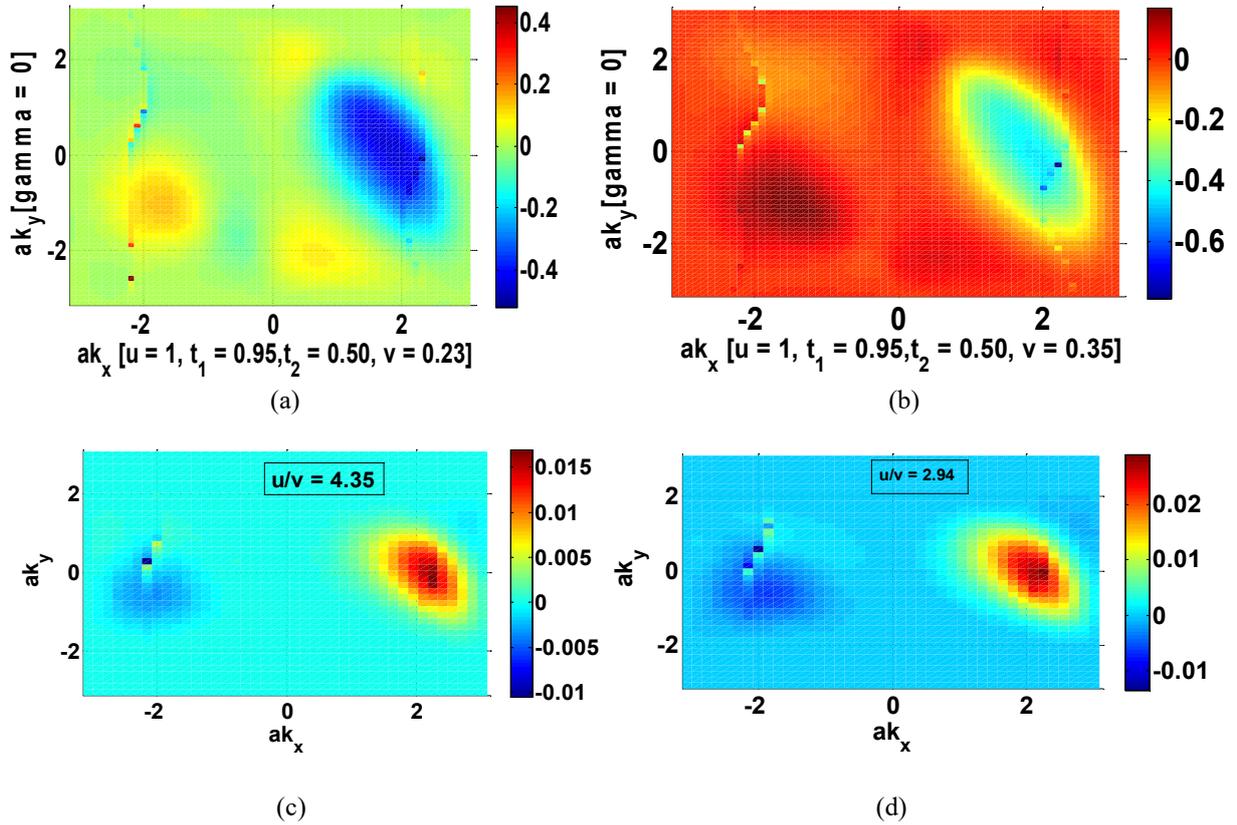

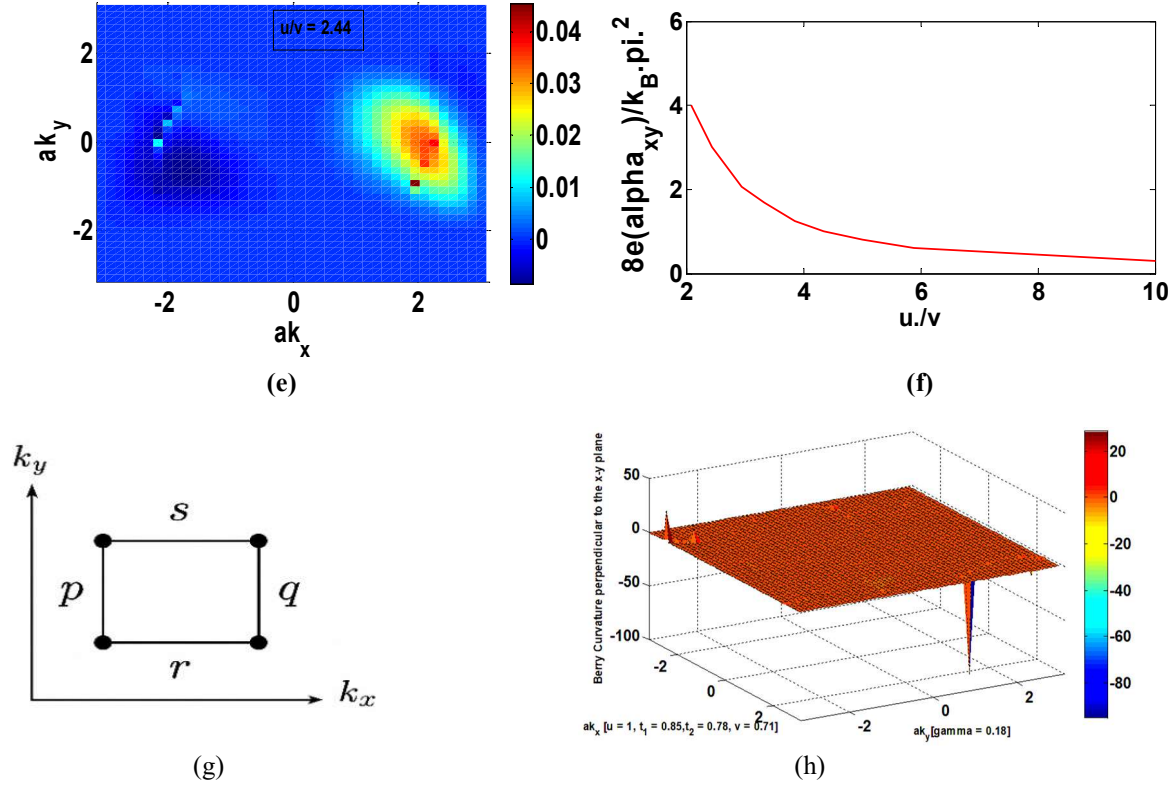

**Figure 6. (a) and (b)** The plots of the Berry curvature in the z-direction as a function of $(ak_x, ak_y)$. The numerical values of the parameters used in the plots are $u = 1, t_1 = 0.95, t_2 = 0.50$, $v = 0.23\,(0.35), \mu = 0,$ and $\gamma = 0$. **(c), (d) and (e)** The contour plots of the integrand $\Omega_{xy}^{(n)} \frac{\partial f(E_n(k))}{\partial E_n(k)}$ in (18). The numerical values of the parameters used in the plots are $u = t_1 = 1, t_2 = 0.5$, $\mu = 0,$ and $\gamma = 0$. The parameter $v$ is 0.23, 0.34, and 0.41 in (c), (d) and (e), respectively. The choice of the values ensures broken IS. **(f)** The plot of the Nernst conductivity $\alpha_{xy}$ as a function of $(u/v)$. **(g)** A path in the reciprocal space showing a plaquette for the application of FHS method. The figure also displays the couplings $(s,r)$ and $(q,p)$. **(h)** A plot of the Berry curvature in the z-direction (obtained by the FHS method) in the staggered hopping amplitude case. The numerical values of the parameters used in the plots are $u = 1, t_1 = 0.85, t_2 = 0.78, v = 0.71,$ and $\gamma = 0.18$.

Here, $\mu$ is the chemical potential, and $f(\varepsilon)$ is the Fermi-Dirac distribution. At non-zero temperatures, however, the formula $\alpha_{xy}(\mu, T) = k_B \frac{e}{\hbar} \sum_n \int d\mathbf{k}\, \Omega_{xy}^{(n)} s(E_n(k))$ where the entropy density $s(E_n(k))$ is given by the expression $s(\varepsilon) = \frac{\varepsilon - \mu}{k_B T} f(\varepsilon) + \log\left(1 + \exp\left(\frac{\mu - \varepsilon}{k_B T}\right)\right), f(\varepsilon) = \frac{1}{1 + \exp\left(\frac{\varepsilon - \mu}{k_B T}\right)}$ where $k_B$ the Boltzmann constant. The integrand of the above expression is contour plotted in Figure 6(c), 6(d) and 6(e) as a function of $(ak_x, ak_y)$ for the different values of $(u/v)$ as indicated in these figures. The numerical values of the parameters used in the plots are $u = t_1 = 1, t_2 = 0.5$, $\mu = 0$, and $\gamma = 0$. The parameter $v$ is 0.23, 0.34, and 0.41 in (c), (d) and (e), respectively. The choice of the values ensures broken IS. The $k-$integration process is facilitated utilizing the Matlab package. The Nernst conductivity $\alpha_{xy}$ is plotted as a function of $(u/v)$ in Figure 6(f). The plot illustrates that as the ratio $(u/v)$ increases, the Nernst response correspondingly decreases.

The expression of the anomalous Hall conductance (AHC) is $\sigma_{AH} = \left(\frac{e^2}{h}\right) \sum_{\alpha \in \text{occupied bands}} \int_{BZ} \frac{d^2k}{(2\pi)^3} f(E_\alpha(k) - \mu) \Omega_\alpha^z(k)$, where $f(E_\alpha(k) - \mu)$ symbolizes the fermion occupation probability of the band $E_\alpha(k)$. Whereas AHC is contributed from the sum of Berry curvatures of all occupied bands, ANE is computed by the Berry curvature of the bands close to the Fermi level. We utilized the Matlab package, as before, to obtain an approximate value of $\sigma_{AH}$ referring to the plots, of the RR-Berry curvature in the z-direction, shown in Figure 6 (a) and (b). We obtain $\sigma_{AH} \approx C\left(\frac{e^2}{\hbar}\right)$ where $C = -2.3235(-2.2290)$ for Figure 6(a) (6(b)). We have not been able to find a suitable selection of the number of grids for which the Chern number ($C = \int_{BZ} \Omega_\alpha^z(k) \frac{d^2k}{(2\pi)^2}$) quantization is possible. We have also explored this non-quantization issue with the parameter values $u = t_1 = 1, t_2 = 0.75$, $v = 0.62, \mu = 0$, and $\gamma = 0.32(0.77)$ leading to TRS and IS disruption. The outcome replicated that of the earlier case. Therefore, notwithstanding the broken IS and/or TRS, anticipated to yield the possibility of the quantum anomalous Hall effect, we find that the Chern number quantization, seemingly, may not be feasible here.

To verify the non-quantization aspect of the Chern number, we have also applied the Fukui-Hatsugai-Suzuki (FHS) method **[57,58]** to our model, assuming a discretized Brillouin zone. This approach leverages lattice gauge theory, employing discrete link variables instead of derivatives. The lattice boundary condition (BC) is gauge-invariant, as it is constructed from a Wilson loop or product of link variables around a plaquette, rendering it invariant under local phase redefinition of the wave functions. Figure 6(g) illustrates the plaquette in the reciprocal space and the couplings $(s,r)$ and $(q,p)$. We employed the following numerical values of the hopping parameters/ISP: $u = 1, t_1 = 0.85, t_2 = 0.78$, $v = 0.71, \mu = 0$, and $\gamma = 0.18$. Figure 6(h) displays the total BC representation across the Brillouin zone, which roughly aligns with our qualitative analysis at the high-symmetry points (HSPs). At the $\Gamma(0,0)$ point, the couplings exhibit finite values, distinct from zero, for the chosen parameter values. The eigenvalues of $\pm 3.3371$ and $\pm 0.3317$, including a balanced and symmetric structure of the eigenvectors, indicate a PT-unbroken feature at this point. The symmetry between eigenvectors corresponding to $\pm$ eigenvalues implies that their Berry curvatures cancel each other, resulting in zero net Berry curvature. However, individual bands retain non-zero Berry curvature.

For the other HSPs, the couplings $(s,r)$ or/and $(q,p)$ are close to zero, which may result in possible BC concentration at these points. At the $X(\pi, 0)$ point, there is eigenvalue degeneracy: 1.4910 and –1.4910 each appear twice. As regards the eigenvectors, we find there is strong symmetry and conjugation; each eigenvector has a "partner" with mirrored real/imaginary parts and signs. Even though eigenvalues are real, the complex structure of eigenvectors obtained suggests that the curvature is likely to peak near point(s) where the degeneracy is lifted. We obtain almost the similar feature at the $Y(0, \pi)$ point, viz. two copies of degeneracy each of $\pm 1.8460$, for each eigenvector, there's a partner with mirrored real and imaginary parts, and the symmetry in eigenvectors and eigenvalues suggests that Berry curvature contributions from $\pm E$ bands may cancel in total, but each band can carry non-zero curvature individually. The $M$ point scenario, however, is different. The eigenvalue matrix is diagonal [–0.1697, + 0.1149i, 0.1697, – 0.1149i]. The structural patterns between eigenvectors with ±imaginary eigenvalues are similar. The curvature might have imaginary components, indicating non-conservative or dissipative topological responses. The computed Chern number ($C$) is approximately ($-2.2222$). No matter whatever be the selection of the number of plaquettes, $C$ results in a non-quantized value. A plausible explanation for the non-quantization is that, for non-Hermitian systems with balanced gain and loss, the corresponding curvature may not be well-behaved, leading to a breakdown in the standard definition of the Chern number.

## 5. Concluding remarks and future perspective

This paper explores a 2D non-Hermitian variant of the SSH model. Our model Hamiltonian's energy eigenvalues are given by the quartic $Q_{A,J,\gamma}(\lambda) = 0$. We find that decreasing the numerical value of the tunable gain/loss parameter $\gamma$ transforms the system from an insulator to a conductor. Notably, when $J = 0$, $Q_{A,\gamma}(\lambda) = 0$ leads to a square-root branch point structure typical of multi-level exceptional point (EP) physics. The utilization of the rank-nullity theorem and graphical analysis of the phase rigidity factor facilitate the identification of such points. Moreover, the Zak phase, as determined in section 3, is an inherently geometric object unaffected by gauge selection and unit cell origin. In our 2D system with unbroken inversion symmetry, we observe that the vectorized Zak phase components, functioning as a topological index for band characterization, display values of either 0 or $\pi$ for $\gamma = 0$ and $\gamma \neq 0$ in our graphical representations in Figure 4. The plots reveal that the conventional bulk-boundary correspondence is highly dependent on the ratio $(u/v)$. These values indicate the existence or non-existence of gapless edge states. Importantly, the inversion symmetry-protected quantized Zak phase by itself is not quite adequate to establish the bulk-boundary correspondence, as edge states may disappear in the topological nontrivial phase, even in Hermitian systems **[59-61]**, resulting in a breakdown of the conventional bulk-boundary correspondence. Furthermore, our analysis of the topolectric RLC circuit reveals that adherence to the robust topological boundary resonance (TBR) condition is attainable in the circuit when specific condition ($R \leq 1\ Ohm$) is fulfilled; a larger value of $R$ does not favor TBR. The quantum Hall susceptance is calculated using the eigenfunctions corresponding to the Laplacian matrix in Eq.(13) above, yielding a positive value indicative of capacitive properties. We have also examined specific topological properties of the model (Eq. (5)) in section 4, finding that staggered hopping amplitudes lead to broken IS with non-zero BC, resulting in finite anomalous Nernst conductivity. However, we find that the Chern number quantization is not possible utilizing RR-BC. We have reconfirmed the Chern number calculation using FHS method **[57,58]** for our non-Hermitian model, and our result remains unchanged. These findings represent the primary highlights of the present work.

The incorporation of asymmetric hopping or gain/loss profiles is one of the mechanisms of the origin of the non-reciprocal effects in non-Hermitian systems. This asymmetry may lead to the non-Hermitian skin effect (NHSE)**[62]**, in which bulk states migrate toward boundaries, creating edge states that dominate transport. In Hermitian systems, edge states can be predicted from bulk invariants, such as winding number. NHSE invalidates this correspondence-bulk behaviour under PBC will not predict the edge behaviour under OBC. The case in point is the 2D non-Hermitian Hatano-Nelson model **[63]**. One of our future goals is to explore this system using non-Bloch theory leading to generalized Brillouin zone (GBZ). We expect our treatment to predict correctly the presence or absence of edge states in the presence of NHSE. To explain the next goal, we pose the question, what occurs when two disparate dimensional SSH systems, each exhibiting a symmetry-protected topological phase in its respective parent dimension, are coupled together? What influence do symmetries exert in such systems? Our aim is to investigate the captivating interplay between a one-dimensional SSH and an environment, characterized by the present two-dimensional SSH system, when coupled together. In general, they are expected to act as boundaries to each other, giving rise to the emergence of different dimensional gapless boundary modes in the same composite system: a zero-dimensional edge mode of the former and one-dimensional chiral boundary states hosted by the latter. Our future aim is to thoroughly investigate these two problems, with the expectation of revealing new perspectives in the study of non-Hermitian physics and the development of non-Hermitian systems.

## Appendix A

The right eigenstates linked to the eigenvalues of (5) could be written down in an explicit manner as $|u^{(j)}(k_x, k_y)\rangle = N_{j0}^{-\frac{1}{2}}\phi_j(k_x, k_y)$, where $\phi_j(k_x, k_y)$ is the transpose of the row vector ($\psi_1^{(j)}(\mathbf{k})$ $\psi_2^{(j)}(\mathbf{k})$ $\psi_3^{(j)}(\mathbf{k})$ $\psi_4^{(j)}(\mathbf{k})$), $j = (1, 2, 3, 4)$, $\mathbf{k} = (k_x, k_y)$. The normalization factor $N_{j0}$ needs to be determined using the bi-orthonormality condition $\langle v^{(i)}|u^{(j)}\rangle = \delta_{ij}$. The elements $\psi_\alpha^{(j)}(k)$ ($\alpha = 1,2,3,4$) are given by $\psi_1^{(j)}(k) = \Delta_{10}^{(j)} + i\Delta_{11}^{(j)}, \psi_2^{(j)}(k) = \Delta_{20}^{(j)} + i\Delta_{21}^{(j)}, \psi_3^{(j)}(k) = \Delta_{30}^{(j)} + i\Delta_{31}^{(j)}$, and $\psi_4^{(j)}(k) = \Delta_{40}^{(j)} + i\Delta_{41}^{(j)}$, where for the $j^{th}$ band

$$\Delta_{10}^{(j)} = (uvt_1 + vt_1^2\cos(ak_x)) + ut_1t_2\cos(ak_y) + t_2t_1^2\cos(ak_x + ak_y) + u^2v\cos(ak_x) +$$
$$uvt_1\cos(2ak_x) + t_2u^2\cos(ak_x + ak_y) + ut_1t_2\cos(2ak_x + ak_y) - (t_2 + v\cos(ak_y)) \times$$
$$\{\lambda_j^2(k) - \gamma^2 - |p|^2\}, \qquad (A.1)$$

$$\Delta_{11}^{(j)} = (vt_1^2\sin(ak_x)) + ut_1t_2\sin(ak_y) + t_2t_1^2\sin(ak_x + ak_y) + u^2v\sin(ak_x) +$$

$$uvt_1\sin(2ak_x) + t_2u^2\sin(ak_x + ak_y) + ut_1 t_2\sin(2ak_x + ak_y) - \left(v\sin(ak_y)\right) \times$$
$$\{\lambda_j^2(k) - \gamma^2 - (v + t_2\exp(iak_y))(v + t_2\exp(-iak_y))\}, \quad (A.2)$$

$$\Delta_{20}^{(j)} = \lambda_j(k)\{\left(v + t_2\cos(ak_y)\right)\left(t_1 + u\cos(ak_x)\right) - u t_2\sin(ak_x)\sin(ak_y) + \left(t_2 + v\cos(ak_y)\right)\left(u + t_1\cos(ak_x)\right) - v t_1\sin(ak_x)\sin(ak_y)\} +$$
$$\gamma\{t_2\sin(ak_y)(t_1 + u\cos(ak_x)) + u\sin(ak_x)\left(v + t_2\cos(ak_y)\right) - v\sin(ak_y) \times$$
$$(u + t_1\cos(ak_x)) - t_1\sin(ak_x)\left(t_2 + v\cos(ak_y)\right)\}, \quad (A.3)$$

$$\Delta_{21}^{(j)} = \lambda_j(k)\{\left(t_2\sin(ak_y)\right)(t_1 + u\cos(ak_x)) + u\sin(ak_x)\left(v + t_2\sin(ak_y)\right) +$$
$$\left(v\sin(ak_y)\right)(u + t_1\cos(ak_x)) - t_1\sin(ak_x)(t_2 + v\cos(ak_y))\} -$$
$$\gamma\{\left(v + t_2\cos(ak_y)\right)(t_1 + u\cos(ak_x)) + u t_2\sin(ak_x)\sin(ak_y) +$$
$$\left(t_2 + v\cos(ak_y)\right)(u + t_1\cos(ak_x)) - v t_1\sin(ak_x)\sin(ak_y)\}, \quad (A.4)$$

$$\Delta_{30}^{(j)} = |s|^2\left(v + t_2\cos(ak_y)\right) - (\lambda_j^2(k) - \gamma^2)(t_1 + u\cos(ak_x)) + (uvt_2 +$$
$$uv^2\cos(ak_y)) + ut_2^2\cos(ak_y) + uvt_2\cos(2ak_y) + t_1t_2v\cos(ak_x) + t_1 \times$$
$$(t_2^2 + v^2)\cos(ak_x + ak_y) + t_1t_2v\cos(2ak_y + ak_x), \quad (A.5)$$

$$\Delta_{31}^{(j)} = -|s|^2\left(t_2\sin(ak_y)\right) + (\lambda_j^2(k) - \gamma^2)(u\sin(ak_x)) - uv^2\sin(ak_y)$$
$$-ut_2^2\sin(ak_y) - uvt_2\sin(2ak_y) - t_1t_2v\sin(ak_x) - t_1(t_2^2 + v^2)\sin(ak_x + ak_y)$$
$$-t_1t_2v\sin(2ak_y + ak_x), \quad (A.6)$$

$$\Delta_{40}^{(j)} = (\lambda_j^3(k) - \lambda_j(k)\gamma^2 - \lambda_j(k)|p|^2 - \lambda_j(k)|s|^2), \quad (A.7)$$

$$\Delta_{41}^{(j)} = \gamma^3 + \gamma|p|^2 + \gamma|s|^2 - \gamma\lambda_j^2(k). \quad (A.8)$$

The quantum geometric tensor is defined in Section 4. The real part of the QGT is the quantum metric and the imaginary part gives BC. As regards the latter, utilizing the Heisenberg equation of motion is $i\hbar \frac{d\hat{x}}{dt} = [\hat{x}, \hat{H}]$, we find that the identity

$$\hbar\langle u^{(\alpha)}(k')|\hat{v}_j|u^{(\alpha)}(k)\rangle = (E_\alpha(k') - E_\alpha(k))\langle u^{(\alpha)}(k')|\frac{\partial}{\partial k_j}|u^{(\alpha)}(k)\rangle \quad (A.9)$$

is satisfied for a system in a periodic potential and its Bloch states as the eigenstates $|u^{(\alpha)}(k)\rangle$. Here the operator $\hbar^{-1}\frac{\partial H(k)}{\partial k_j} = \hat{v}_j$ represents the velocity in the $j = (x,y)$ direction. Upon using the identity above, the z-component of BC may now be written as

$$\Omega_{xy}(k) = -2\sum_\alpha Im\left\langle\frac{\partial u^{(\alpha)}(k)}{\partial k_x}\bigg|\frac{\partial u^{(\alpha)}(k)}{\partial k_y}\right\rangle. \quad (A.10)$$

We use this formula to present the outline of the calculation of BC below. It is not difficult to see that for the present problem the product

$$\left\langle \frac{\partial u^{(\alpha)}(k)}{\partial k_x} \bigg| \frac{\partial u^{(\alpha)}(k)}{\partial k_y} \right\rangle = \Sigma_{j=1,2,3,4} \left[ \left( P_{jx}^{(\alpha)} P_{jy}^{(\alpha)} + Q_{jx}^{(\alpha)} Q_{jy}^{(\alpha)} \right) + i \left( P_{jx}^{(\alpha)} Q_{jy}^{(\alpha)} - Q_{jx}^{(\alpha)} P_{jy}^{(\alpha)} \right) \right], \quad (A.11)$$

where

$$P_{jx}^{(\alpha)} = -\left(\frac{1}{2}\right) N_\alpha^{-\frac{3}{2}} (\partial_x N_\alpha) \Delta_{j0}^{(\alpha)} + N_\alpha^{-\frac{1}{2}} (\partial_x \Delta_{j0}^{(\alpha)}), Q_{jy}^{(\alpha)} = -\left(\frac{1}{2}\right) N_\alpha^{-\frac{3}{2}} (\partial_y N_\alpha) \Delta_{j1}^{(\alpha)} + N_\alpha^{-\frac{1}{2}} (\partial_y \Delta_{j1}^{(\alpha)}),$$

(A.12)

$$Q_{jx}^{(\alpha)} = -\left(\frac{1}{2}\right) N_\alpha^{-\frac{3}{2}} (\partial_x N_\alpha) \Delta_{j1}^{(\alpha)} + N_\alpha^{-\frac{1}{2}} (\partial_x \Delta_{j1}^{(\alpha)}), P_{jy}^{(\alpha)} = -\left(\frac{1}{2}\right) N_\alpha^{-\frac{3}{2}} (\partial_y N_\alpha) \Delta_{j0}^{(\alpha)} + N_\alpha^{-\frac{1}{2}} (\partial_x \Delta_{j0}^{(\alpha)}),$$

(A.13)

$$(\partial_{x/y} N_\alpha) = 2 \Sigma_{j=1,2,3,4} [\Delta_{j0}^{(\alpha)} \left( \partial_{x/y} \Delta_{j0}^{(\alpha)} \right) + \Delta_{j1}^{(\alpha)} (\partial_{\frac{x}{y}} \Delta_{j1}^{(\alpha)})]. \quad (A.14)$$

The symbol $\partial_x$ ($\partial_y$) above stands for the differential coefficient $\frac{\partial}{\partial k_x}$ ($\frac{\partial}{\partial k_y}$). Now that we have calculated a formal expression for the BC of α band, what remains to be done is to calculate various derivatives in (16).